# Fieldable Muon Momentum Measurement using Coupled Pressurized Gaseous Cherenkov Detectors

J. Bae*, S. Chatzidakis*

*School of Nuclear Engineering, Purdue University, West Lafayette, IN 47906, bae43@purdue.edu
*doi.org/10.13182/T125-36581*

## INTRODUCTION

Cosmic ray muons present a large part of the radiation background and depending on the application of interest muons can be seen as background noise, e.g., radiation mapping, radiation protection, dosimetry, or as a useful interrogation probe such as cosmic ray muon tomography. It is worth noting recent developments on muon scattering tomography which has emerged as a prospective non-invasive monitoring method for many applications including spent nuclear fuel cask monitoring [1]–[5] and geotomography [6]. However, it is still very challenging to measure muon momentum in the field, despite the apparent benefits [7], without resorting to large and expensive calorimeters, ring imagers, or time of flight detectors. Recent efforts at CNL and INFN have developed large prototypes based on multiple Coulomb scattering coupled with the muon momentum reconstruction algorithms [8], [9]. While these efforts show promise, no portable detectors exist that can measure muon momentum in the field.

In this work, we present a new concept for measuring muon momentum using coupled pressurized gaseous Cherenkov radiators. By carefully selecting the gas pressure at each radiator we can optimize the muon momentum threshold for which a muon signal will be detected. This way, a muon passing through the radiators will only trigger those radiators with momentum threshold less than the actual muon momentum. By measuring the presence of Cherenkov signals in each radiator, our system can then estimate the muon momentum. The primary benefit of such a concept is that it can be compact and portable enough so that it can be deployed in the field separately or in combination with existing tomography systems.

## MOMENTUM MEASUREMENT USING CHERENKOV RADIATION

A well-known way to measure muon momentum is by measuring the amount of Cherenkov radiation produced when a muon travels through a radiator. However, a muon must have sufficient energy to exceed the threshold momentum level of the radiator so that a sufficiently strong Cherenkov radiation signal is generated. The threshold momentum level depends the refractive index of radiator, $n$ and it is given by:

$$p_{th}c = \frac{m_\mu c^2}{\sqrt{n^2 - 1}} \quad (1)$$

where $m_\mu c^2 = 105.658$ MeV. Molecular refractivity, $A$ in m$^3$/mol, can be described with the molecular polarizability, $\alpha$ in $10^{-30}$ m$^3$:

$$A = \frac{4\pi}{3} N_A \alpha \quad (2)$$

where $N_A$ is the Avogadro number. For gaseous radiators, $n^2 \approx 1$, the refractive index depends on pressure and temperature. It can be estimated by substituting Eq. (2) into the Lorentz-Lorenz equation [10]:

$$n \approx \sqrt{1 + \frac{3Ap}{RT}} \quad (3)$$

where $R$ is the universal gas constant which has units of J/mol-K, $p$ and $T$ are pressure in Pa and temperature in K, respectively. By varying the gas pressure or temperature, the muon momentum threshold levels can be selected.

## PHOTON SIGNALS

Under certain conditions, Cherenkov and scintillation photons are produced when muons travel through a medium. Cherenkov and scintillation photons are in the visible and ultraviolet region of the electromagnetic spectrum. However, scintillation photons are isotropic and have a peak in their frequency spectra, whereas Cherenkov photons are directional and do not have a characteristic spectral peak. We use this observation to separate Cherenkov from scintillation photon signals as muons travel through our detector set up.

The Cherenkov photon yield is proportional to the photon wavelength with the most probable wavelength in the range of visible and ultraviolet (200 to 700 nm). Theoretical Cherenkov photon yield per distance, $dN_{ch}/dx$ can be estimated using the Frank and Tamm equation [11]:

$$\frac{dN_{ch}}{dx} = 2\pi\alpha \left(\frac{1}{\lambda_2} - \frac{1}{\lambda_1}\right)\left(1 - \frac{1}{n^2\beta^2}\right) \quad (4)$$

where $\lambda_1, \lambda_2$ are upper and lower wavelength limits and $\alpha$ is the fine structure constant.

The scintillation photon yield is proportional to the incident particle energy loss per unit distance. Using the Bethe-Bloch equation to calculate the energy loss, we can estimate the scintillation photon yield from [12]:

$$\frac{dN_{sc}}{dx} = S \frac{dE/dx}{1 + k_B(dE/dx)} \quad (5)$$

where $S$ is scintillation efficiency and $k_B$ is Birks' constant. For $E \geq 100$ keV, $k_B \approx 0$.

## COUPLED PRESSURIZED GASEOUS CHERENKOV RADIATORS

To develop a fieldable muon momentum detection system we propose the use of coupled pressurized gaseous Cherenkov radiators with different refractive indices. The refractive index and by extension the muon momentum threshold for each radiator is carefully selected, by varying the gas pressure in each radiator. Different gases can be used to achieve the desired muon momentum threshold. For example, $CO_2$ and $C_4F_{10}$ (Perfluorobutane) are well-known Cherenkov radiators. They are not reactive and have relatively high refractive indices. $CO_2$ is used as a momentum threshold Cherenkov counter for electrons and hadrons at Thomas Jefferson National Accelerator Facility [13], and $C_4F_{10}$ is used in the RICH detector at the LHCb experiment as one of two Cherenkov radiators at CERN. Dry air is not a typical Cherenkov radiator due to its high scintillation photon yield, however it is included in this study for comparison. Their properties are summarized in TABLE 1. Fig 1 shows the variation of muon momentum threshold and refractive indices for $CO_2$, $C_4F_{10}$, and dry air using Eq. (1) and (3) as a function of pressure. The refractive index is linear in this pressure range (< 10 atm) because the dielectric susceptibility, $\chi = n^2 - 1$ is a linear function of pressure with a slope of $3A/RT$ in Eq. (3). The threshold momentum level slowly changes at the high-pressure range.

In our proposed design, we use a combination of twelve radiators. The radiators numbers are assigned from #0 to #11 with a momentum interval of 0.5 GeV/c ranging from 0.1 to 5 GeV/c. The first and last radiators (#0 and #11) are auxiliary radiators designed for lower (< 0.1 GeV/c) and upper (> 5.0 GeV/c) momentum cutoffs measurement. Additionally, there are ten active radiators, in the middle, two solid and eight gaseous Cherenkov radiators are placed in series along the incident muon path. Two scintillator panels are installed on the top and bottom of the pile of radiators for background signal discrimination. Each radiator is designed to be triggered when a muon has at least momentum higher than a pre-selected level in that particular radiator. In other words, depending on the muon momentum, none to all Cherenkov radiators can be triggered. For the two lowest active momentum levels ($p_{th} = 0.1$ and 0.5 GeV/c), we use solid radiators (glass) to achieve the necessary refractive indices that will allow a very low momentum threshold not possible with a gas radiator. The rest of the eight active radiators use gas at different pressure levels. Fig 2 shows an example for an incident muon with energy 4.1 GeV. #0 to #9 bins are triggered because the corresponding muon momentum is larger than 4.0 GeV/c. In this case, the system records '1' for triggered bins regardless of its signal amplitude and '0' for the non-triggered bins. Following differentiation and signal analysis, the final signal indicates the muon momentum range. At each radiator we use wavelength shifting fibers (WSFs) on the bottom and photon absorber liners. WSFs convert UV lights to visible lights and guide them to photon counters to convert photon signals to electrical signals. The photon absorber liners prevent photons from escaping and suppress the scintillation photon signals given that scintillation photons are emitted in all directions whereas Cherenkov photon emission is directional and only detected from the bottom of the chamber. Fig 3. shows the corresponding circuit block. A digital signal analyzer accepts the signal from two inputs, one from the coincidence logic gate to operate signal analysis, the other from twelve photon count signals from each Cherenkov radiator.

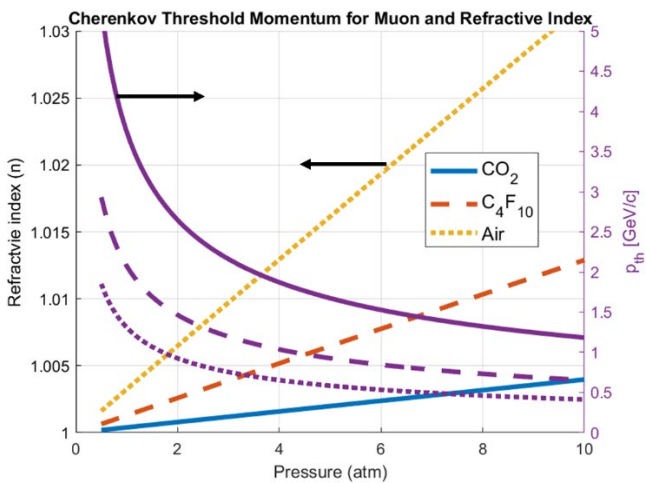

Fig 1. Cherenkov muon momentum threshold level and refractive index for $CO_2$, $C_4F_{10}$, and air as a function of pressure.

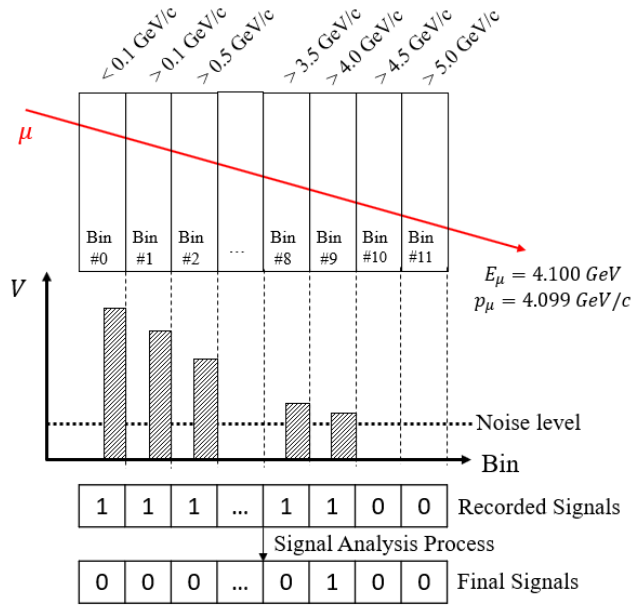

Fig 2. Schematic showing the conceptual process for identifying muon momentum when the Cherenkov radiators are triggered by a muon with an energy 4.1 GeV.

TABLE I. Properties of gaseous Cherenkov radiators

|  | $CO_2$ | $C_4F_{10}$ | Air |
|---|---|---|---|
| Molecular Weight (g/mol) | 44.01 | 238.03 | 28.96 |
| Molecular Polarizability, $\alpha$ ($10^{-30}$m$^3$) [14], [15] | $2.59 \pm 0.01$ | $8.44 \pm 0.12$ | $21.18 \pm 0.91$ |
| Refractive Index (at STP), $n$ [16], [17] | 1.00045 | 1.0014 | 1.000273 |
| Scintillation Efficiency, $S$ (photons/MeV) (UV+VIS) [18], [19] | $5.09 \pm 0.28$ | $3.1 \pm 1.6$ | $25.46 \pm 0.43$ |

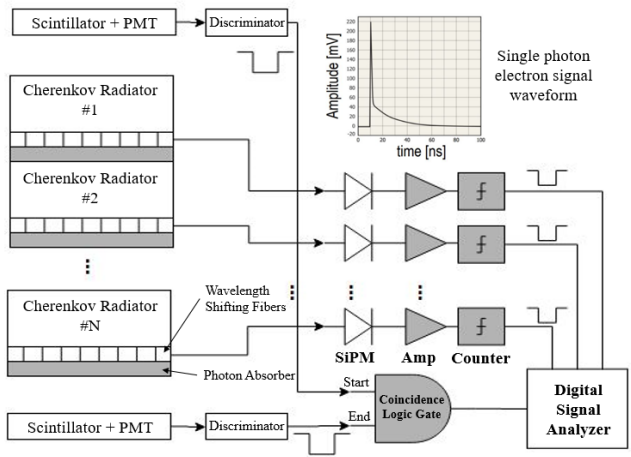

Fig 3. Circuit block of coupled pressurized gaseous Cherenkov radiator muon momentum detection system.

**RESULTS**

Three gaseous Cherenkov radiators, $CO_2$, $C_4F_{10}$, and dry air are simulated using Monte Carlo to evaluate their Cherenkov characteristics and identify the optimal radiator for muon momentum measurement. Both Cherenkov and scintillation photon yields depend on the traveling distance of the particle. A larger in size radiator would produce more Cherenkov photons, but also more noise from scintillation photons. In addition, a larger radiator would increase the absorption probability of photons thereby reducing the available photons. To better understand the effect of radiator size, two different scenarios are studied, (i) one with uniformly sized radiators with 10 cm width, and (ii) the other with diverse sized radiators with 0.5 cm for solid radiators, 10 cm for high- and 20 cm for low-pressure gas radiators. For each scenario, we simulated $10^4$ muons with uniformly distributed energies from 1 to 5.5 GeV and measured the number of muons that their momentum was correctly predicted. Fig 4 shows the $C_4F_{10}$ classification rates per muon momentum for each scenario. Classification rate is defined as the probability that a system correctly categorizes the incident muon momentum. Muon momentum is labeled as average momentum because each bin represents the momentum range with an interval of 0.5 GeV/c from 0.1 to 5.0 GeV/c. To suppress noise, a discriminator is used, which deducts a single signal for all inputs. $CO_2$ and $C_4F_{10}$ with a discriminator estimate the muon momentum with high classification rate (~90%) at most momentum levels. The average classification rate over all energies for uniform radiator size for $CO_2$, $C_4F_{10}$, and dry air is 69.5, 70.0, and 48.7 %. The average classification rate for diverse radiator size for $CO_2$, $C_4F_{10}$, and dry air is 83.5, 83.5, and 56.0 %. Average classification rates for $CO_2$, $C_4F_{10}$, and air with four scenarios are also shown in Fig. 5. From this, we conclude $C_4F_{10}$ and $CO_2$ with a diverse size perform equally well at identifying muon momentum.

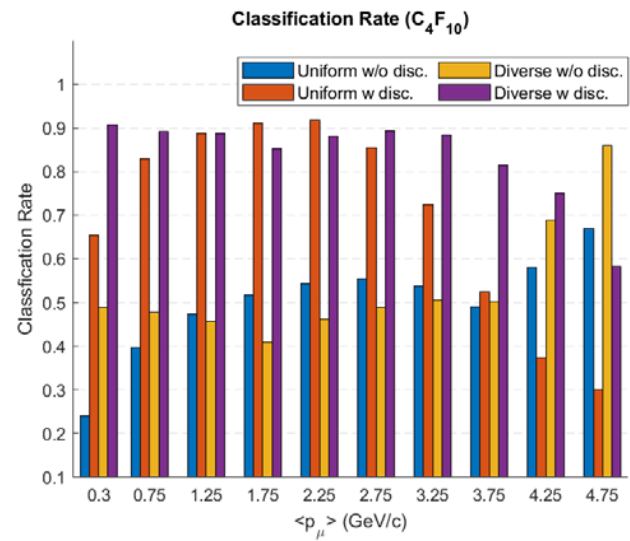

Fig 4. $C_4F_{10}$ classification rates for average muon momenta from 0.3 to 4.75 GeV/c for four scenarios.

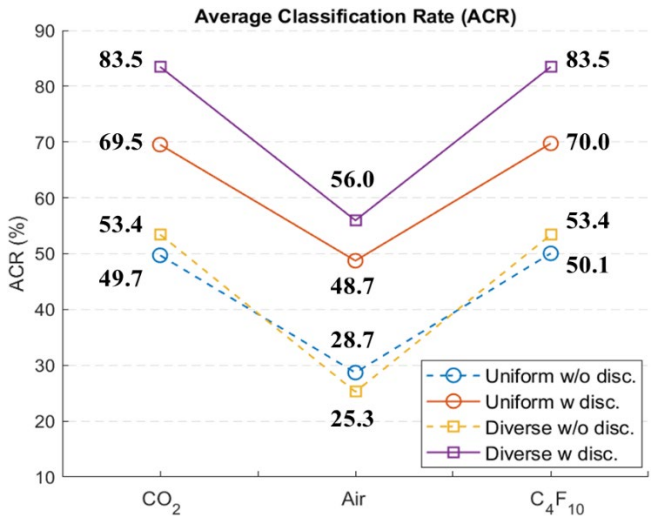

Fig 5. Average classification rates of $CO_2$, $C_4F_{10}$, and air for four scenarios.

**CONCLUSIONS**

In this work, we explored a new concept for muon momentum measurement using Cherenkov radiation from multiple pressurized gaseous Cherenkov radiators. The goal is to develop a fieldable prototype detector that can be used to measure the muon momentum for field applications such as background radiation mapping, radiation protection, muon tomography, nuclear nonproliferation, and reactor monitoring. We simulated muon transport through three gaseous Cherenkov radiators, $CO_2$, $C_4F_{10}$, and air using Monte Carlo and evaluated their responses. $C_4F_{10}$ and $CO_2$ outperform dry air, successfully estimating the average incoming muon momentum with resolution, $\sigma_p = 0.5$ GeV/c and with a high classification rate (~90%). There do exist additional factors yet to be considered that will decrease the classification rate. Measurement errors, less than perfect detector efficiency, photons production other than Cherenkov and scintillation, various noise sources in electronics, and more complicated geometries are real-world considerations that will broaden the distributions. However, the results presented herein motivate a more detailed study in order to identify its performance under real world conditions.

**ACKNOWLEDGEMENTS**

This research is being performed using funding from the Purdue College of Engineering and the School of Nuclear Engineering.